\documentclass[12pt]{article}
\usepackage[utf8]{inputenc}
\usepackage[T1]{fontenc}
\usepackage{graphicx}
\usepackage{longtable}
\usepackage{wrapfig}
\usepackage{rotating}
\usepackage[normalem]{ulem}
\usepackage{amsmath}
\usepackage{amssymb}
\usepackage{capt-of}
\usepackage{hyperref}
\usepackage[margin=1in]{geometry}
\usepackage{mathrsfs}
\usepackage{bbold}
\usepackage{cancel}
\usepackage{braket}
\usepackage{authblk}

\usepackage{setspace}
\usepackage{xcolor}
\usepackage{amsmath,amsfonts,epsfig}
\usepackage{tcolorbox}
\usepackage{cite}
\usepackage{subcaption}
\usepackage{xstring}
\usepackage[hang,flushmargin]{footmisc} 
\let\svthefootnote\thefootnote
\newcommand\freefootnote[1]{
  \let\thefootnote\relax
  \footnotetext{#1}
  \let\thefootnote\svthefootnote} % for emails

\DeclareMathOperator{\tr}{Tr}
\newcommand\GM[1][]{{\rm GM}^{\IfInteger{#1}{(#1)}{\mathtt{(#1)}}}}
\renewcommand\S[1][]{S^{\IfInteger{#1}{(#1)}{\mathtt{(#1)}}}}

\title{\bf Why many-partite entanglement is essential \\
for holography }
\author[$\clubsuit$,$\diamondsuit$]{Norihiro Iizuka}
\author[$\heartsuit$]{Simon Lin}  
\author[$\spadesuit$]{Mitsuhiro Nishida}
 
\affil[$\clubsuit$]{\it Department of Physics, National Tsing Hua University, Hsinchu 300044, Taiwan}
\affil[$\diamondsuit$]{\it Yukawa Institute for Theoretical Physics, Kyoto University, Kyoto 606-8502, Japan}
\affil[$\heartsuit$]{\it New York University Abu Dhabi, P.O. Box 129188, Abu Dhabi, United Arab Emirates}
\affil[$\spadesuit$]{\it National Institute of Technology, Yuge College, Ehime 794-2593, Japan}

\date{\today}

\begin{document}

\maketitle

\thispagestyle{empty}
\setcounter{page}{0} % start page counting after the title page

\vspace{-.3in}
\begin{abstract}
{We argue that many-partite entanglement is ubiquitous in holography and holographic quantum error correction codes. We base our claim on genuine multi-entropy, a new measure for multi-partite entanglement. We also discuss a connection between the bulk IR reconstruction and many-partite entanglement on a large number of boundary subregions.  
} 
\vskip 0.2in
\centering
\noindent {\it Essay written for the Gravity Research Foundation \\ 2025 Awards for Essays on Gravitation}

\freefootnote{E-mail: 
\href{mailto:iizuka@phys.nthu.edu.tw}{\texttt{iizuka@phys.nthu.edu.tw}}, \href{mailto:simonlin@nyu.edu}{\texttt{simonlin@nyu.edu}}, \href{mailto:mnishida124@gmail.com}{\texttt{mnishida124@gmail.com}}}
\end{abstract}

\setcounter{footnote}{0}

\newpage

\section{Introduction}
\label{sec:intro}
Entanglement and quantum information play a pivotal role in our modern understanding of quantum gravity and AdS/CFT. A groundbreaking progress along the line was marked by the discovery of Ryu-Takayanagi (RT) formula \cite{Ryu:2006bv,Ryu:2006ef}, relating entanglement entropy in the boundary CFT to the area of minimal surfaces in the bulk AdS. This sparked the beginning of ``geometry from entanglement'' program \cite{VanRaamsdonk:2010pw,Maldacena:2013xja}, which established the now common belief that bulk geometrical data is encoded in the entanglement structure of the boundary field theory.

The RT formula tells us that holographic CFT states must possess large entanglement in order to support a bulk geometry. A natural question to ask is: Can we classify the amount of different classes of entanglement of holographic CFT states? Or stated differently, how do entanglement between different number of parties contribute to this picture? Unfortunately, this question cannot be answered just by examining the entanglement entropy itself, as it is only sensitive to {\emph{bipartite}} entanglement in a pure state. 
In terms of tripartite entanglement, Ref.~\cite{Akers:2019gcv} answered this question by studying the reflected entropy \cite{Dutta:2019gen}
and Markov gap \cite{Hayden:2021gno}.\footnote{See also \cite{Mori:2024gwe,Mori:2025mcd} for recent progress and discussions on this topic.} 
However for higher partite entanglements this remains an open problem.

In this essay, we would like to propose an answer to this question by arguing that holographic CFT states must contain large amounts of $\mathtt{q}$-partite entanglement for all\footnote{Here we mean all $\mathtt{q}$ up to ${\cal{O}}\left({V/\ell_{\rm pl}^d}\right)$, where $\ell_{\rm pl}$ is the Planck length scale and $V$ is the $d$-dimensional volume of the boundary CFT.} $\mathtt{q}\ge3$. We base our proposal on two observations: The first being the structure of holographic error correction codes  \cite{Almheiri:2014lwa}, where we argue that under conditions imposed by holography, its code subspace naturally possesses large amounts of multipartite entanglement. The second being a new measure of multipartite entanglement called \emph{genuine multi-entropy} \cite{Iizuka:2025ioc}, which we argue to be non-zero in holographic states from its proposed geometric dual. We briefly review them in the remainder of this section.

\subsection*{Holographic quantum error correction code}
\label{sec:QEC}
An important piece in our modern understanding of the holographic dictionary is the realization of quantum error correction (QEC) in AdS/CFT \cite{Almheiri:2014lwa}. 
The central idea is that one should think of the relation between the bulk and the boundary as a linear map encoding semiclassical bulk operators into the CFT Hilbert space, which can then be ``reconstructed'' from the boundary data \cite{Hamilton:2006az}.

\begin{figure}[ht]
    \centering
    \includegraphics[width=0.9\linewidth]{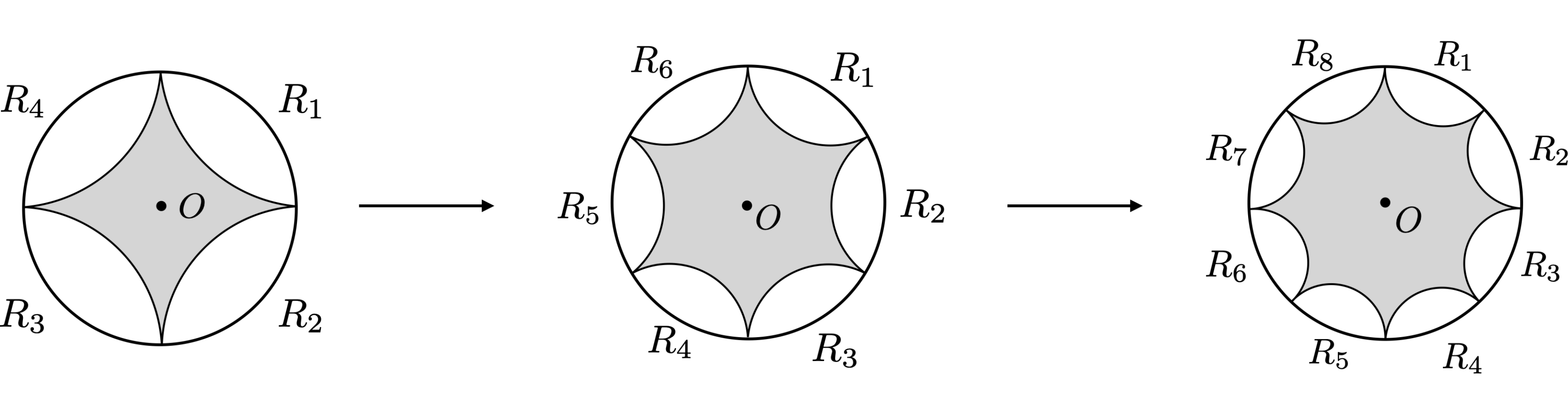}
    \caption{\onehalfspacing The boundary subregions $R_1,R_2,\cdots$ and their corresponding entanglement wedges (shown in white). $O$ stands for an local operator that can be reconstructed even if one loses some subregions. As we divide the boundary into finer subregions, the possible location of $O$ (shown in gray) grows in size. In the boundary field theory, all subregions have independent degrees of freedom.}
    \label{fig:div4}
\end{figure}

As an extension of the causal wedge, the \emph{entanglement wedge} of a boundary subregion $R$ is defined as the bulk region bound by $R$ itself and its RT surface, which has been suggested as the region reconstructible given data on $R$ \cite{Czech:2012bh,Dong:2016eik,Cotler:2017erl,Chen:2019gbt}.  
A salient feature of this entanglement wedge reconstruction is that one does not need access to the entire boundary in order to perform reconstructions. Consequently, a bulk operator can still be reconstructed even if some boundary subregions are lost. Moreover, the selection of the lost subregions is arbitrary, although the number of subregions to be lost is limited depending on the position of reconstructed local operator.
For example, in Fig.~\ref{fig:div4}, the reconstruction of the bulk local operator $O$ is possible even if one loses subregions $R_1$, or $R_2$, etc.

It is long understood that these properties of QEC are only possible due to the entanglement between the subsystems. However, the exact structure these entanglement remains largely unknown. This is one of our main motivation to study $\texttt{q}$-partite entanglement for holographic QECs.

\subsection*{Genuine multi-entropy}
To study $\mathtt{q}$-partite entanglements, we will make use of the multi-entropy \cite{Gadde:2022cqi,Penington:2022dhr,Gadde:2023zzj}, which is a natural generalization of 
entanglement entropy into $\mathtt{q}$-partite subsystems. The multi-entropy, denoted $\S[q]$, is defined as  a logarithm of a symmetric contraction of pure density matrices.  
The multi-entropies can be used to quantify higher-partite entanglements. However, $\S[q]$ itself is not a great measure of $\texttt{q}$-partite entanglement since it is also sensitive to all $\mathtt{\tilde{q}}$-partite entanglements for $\mathtt{\tilde{q}}<\mathtt{q}$. 
Thus, one cannot determine whether the nonzero value of $\S[q]$ is due to bipartite or multipartite entanglement. To resolve this issue, one can define a \emph{genuine multi-entropy} (denoted $\GM[q]$) with the following properties \cite{Iizuka:2025ioc}:
\begin{itemize}
    \item $\GM[q]$ includes the $\mathtt{q}$-partite multi-entropy.
    \item $\GM[q]$ vanishes for all $\mathtt{\tilde{q}}$-partite entangled states with $\mathtt{\tilde{q}} < \mathtt{q}$.\footnote{Example of these states includes but is not limited to separable states. For example, in $\texttt{q}=3$ they also include \emph{triangle states}. They are bipartitely-entangled states of the form $\ket{\psi}=\ket{\psi_{A_1B_2}}\ket{\psi_{B_1C_2}}\ket{\psi_{C_1A_2}}$.}
\end{itemize}
In a nutshell, one should think of $\GM[q]$ as an ``irreducible representation decomposition'' of the different numbers of multi-partite entanglements contained in the multi-entropy. 

In practice, $\GM[q]$ can be explicitly constructed from linear combinations of all the lower-partite $\mathtt{\tilde{q}}$-multi-entropies with $\mathtt{\tilde{q}} \le \mathtt{q}$ \cite{Iizuka:2025ioc,upcoming}. 
For example, the $\mathtt{q}=3$ genuine multi-entropy is given by
\begin{equation}
  \label{GM3}  
  \GM[3](A:B:C) = \S[3](A:B:C)-\frac{1}{2}\left(S(A)+S(B)+S(C)\right).
\end{equation}
For higher partite cases, genuine multi-entropies come with free parameters. This is related to the fact that there are in general many different classes of multi-partite entanglement even for the same $\mathtt{q}$. For $\mathtt{q}= 4$ there is one free parameter $a$:
 \begin{align}
    \label{GM4}
    &\GM[4](A:B:C:D) = \S[4](A:B:C:D) - \frac{1}{3}\left( \S[3](AB:C:D) + \S[3](AC:B:D) + \cdots \right) \nonumber \\
                    &\quad + a\left(S(AB)+S(AC)+S(AD)\right) + (1/3-a)\left(S(A)+S(B)+S(C)+S(D)\right),
\end{align}
where $+\cdots$ stands for summing over all possible permutations of the boundary subregions (total 6 terms). For $\mathtt{q}= 5$ we also have a free parameter $b$:
\begin{align}
    &\GM[5](A:B:C:D:E) = \S[5](A:B:C:D:E)
    -\frac{1}{4}\left(\S[4](AB:C:D:E)+\cdots\right) \nonumber \\
    &\quad + \frac{1+4b}{10}\left(\S[3](AB:CD:E)+\cdots\right)
    +\frac{1-16b}{20}\left(\S[3](ABC:D:E)+\cdots\right) \\
    &\quad -\frac{1+4b}{20}\left(S(AB)+S(AC)+\cdots\right)
    +b\left(S(A)+S(B)+S(C)+S(D)+S(E)\right). \nonumber
\end{align}

We emphasize that the (genuine) multi-entropy can be obtained for any positive integer $\mathtt{q}$ in a straightforward manner even for $\mathtt{q}> 5$, which is a major advantage over other previously proposed multipartite entanglement measures. 

$\vspace{-10mm}$
\section{Code subspace in QEC and multipartite entanglement}
\label{sec:example}

We demonstrate the importance of multipartite entanglement in holography from the viewpoint of QEC. A deep connection between QEC and entanglement has long been discussed \cite{Bennett:1996gf,Knill:1997mom}. The main idea is to encode the original states into a code subspace of the physical Hilbert space with a larger dimension. The encoded states in general have  entanglement between subsystems in the physical Hilbert space, and this  entanglement protects against a quantum error. 
The type of QEC relevant for holography is known as \emph{erasure codes}, in which the error is simply erasing a subsystem in the physical space.

Here for our purpose, we show a simple example where we encode one qubit into four \cite{Grassl:1996eh}. The encoding map is given by
\begin{align}
\label{zerobaronebar}
  \begin{split}
    \ket{0} &\to \frac{1}{\sqrt{2}} \left(\ket{0000}+\ket{1111}\right) \equiv \ket{\bar{0}}, \;\;\;
    \ket{1} \to \frac{1}{\sqrt{2}} \left(\ket{1010}+\ket{0101}\right) \equiv \ket{\bar{1}}.
  \end{split}
\end{align}
This code protects against the erasure of any single qubit in the physical Hilbert space. However upon losing two or more qubits, the information about the original state is lost. The reader is encouraged to regard it as a discretized example of the holographic QEC in Fig.~\ref{fig:div4}, where a single bulk qubit is encoded into the boundary of four subregions.

To get a gist of how error correction works, suppose that the first qubit in the physical Hilbert space was erased. This is equivalent to performing a partial trace $\tr_1$ on the first qubit. Under such an operation, we have
\begin{align}
  \begin{split}
  \ket{\bar{0}}\bra{\bar{0}} &\underset{\tr_{1}}{\longrightarrow} \frac{1}{2} \left(\ket{000}\bra{000}+\ket{111}\bra{111}\right), \;\;\;
  \ket{\bar{1}}\bra{\bar{1}} \underset{\tr_{1}}{\longrightarrow} \frac{1}{2} \left(\ket{010}\bra{010}+\ket{101}\bra{101}\right), \\
  \ket{\bar{1}}\bra{\bar{0}} &\underset{\tr_{1}}{\longrightarrow} \frac{1}{2} \left(\ket{010}\bra{111}+\ket{101}\bra{000}\right), \;\;\;
  \ket{\bar{0}}\bra{\bar{1}} \underset{\tr_{1}}{\longrightarrow} \frac{1}{2} \left(\ket{111}\bra{010}+\ket{000}\bra{101}\right).                               
  \end{split}
\end{align}
It is easy to see that after the operation, the matrix elements in the physical subspace do not mix and thus one can easily read out the original state. The erasure for other qubits in the physical Hilbert space follows a similar story. On the other hand, if two qubits are traced out, we lose the distinguishability. For example, 
if the first and the third are traced out, it is obvious that 
$\ket{\bar{0}}$ and $\ket{\bar{1}}$ become the same. 
If the first and the second are traced out, both $\ket{\bar{1}}\bra{\bar{0}}$ and $\ket{\bar{0}}\bra{\bar{1}}$ vanish and they lose distinguishability. Similarly, erasure of any other two qubits results in the lost of the distinguishability.

A noble property of this example is that all of the encoded states, given by \eqref{zerobaronebar}, have the special form of ``generalized GHZ states'' $\ket{\rm GHZ_\texttt{q}}$ with $\mathtt{q}=4$. It is known that the generalized GHZ state in $\mathtt{q}$-qubits is maximally entangled in some aspects \cite{Gisin:1998ze}. Indeed, one can verify this by explicitly calculating $\mathtt{q}$-partite genuine multi-entropy ${\rm GM}^{(\mathtt{q})}$ for the GHZ state. One can confirm that, for $\mathtt{q}=4$, \cite{upcoming} 
\begin{align}
  {\rm GM}^{(4)}_{\ket{\rm GHZ_4}} &= (1/3-a)\log 2 \,.
\end{align}
Thus the encoded states have nonzero multipartite entanglement in general.

We give another example for five qubits here, commonly known as the \emph{five-qubit perfect tensor code} \cite{Laflamme:1996iw}.
This code protects against the erasure of up to any two qubits in the physical Hilbert space.
The code words of this code are
\begin{align}
    &\ket{\bar{0}} = \frac{1}{4}\Big(\ket{00000}+\ket{10010}+\ket{01001}+\ket{10100}+\ket{01010}-\ket{11011}-\ket{00110}-\ket{11000}\nonumber\\
    & \quad -\ket{11101}-\ket{00011}-\ket{11110}-\ket{01111}-\ket{10001}-\ket{01100}-\ket{10111}+\ket{00101}\Big), \nonumber\\
    &\ket{\bar{1}} = \frac{1}{4}\Big(\ket{11111}+\ket{01101}+\ket{10110}+\ket{01011}+\ket{10101}-\ket{00100}-\ket{11001}-\ket{00111}\nonumber\\
    & \quad -\ket{00010}-\ket{11100}-\ket{00001}-\ket{10000}-\ket{01110}-\ket{10011}-\ket{01000}+\ket{11010}\Big).
\end{align}

Contrary to $\mathtt{q}=4$, these encoded states are not of generalized GHZ types. 
One can also confirm that $\mathtt{q}=5$ genuine multi-entropy\footnote{Precisely speaking the R\'enyi version of the genuine multi-entropy is nonzero in general.} for these states are nonzero in general \cite{upcoming}.
In general, one expects for higher $\mathtt{q} \ge 6$, the encoded states are similar to this $\mathtt{q}=5$ example.
 Similarly, one can expect that the encoded states exhibit a nonzero genuine multi-entropy $\mbox{GM}^{(\mathtt{q})}$ in general:
 \begin{align}
 \GM[q] \neq 0 \qquad \text{(for encoded states)}.
 \end{align}

One might object that we are cherry picking amongst all the available QEC codes for examples where there is large multipartite entanglement in the code subspace, and indeed there are known codes which only require parametrically small entanglement \cite{Bravyi:2024ljv}. However, we argue here that they do not have the desired property to qualify as a \emph{holographic code}. Holographic QEC codes have the very special property that if we consider an operator $O$ very deep in the bulk, then $O$ is protected up to the erasure of one half of the total boundary subsystems -- which happens to be the best one can do without violating the no-cloning principle. 
It has been argued that such codes are only possible with nearly maximal amount of entanglement present \cite{Grassl:2003cdn,2020Quant...4..284H}, of which both of the codes we presented in this section are great examples. Whether maximal entanglement implies large multi-partite entanglement is still an open question. However, based on the examplary codes given in this section and various evidences from holography \footnote{For example see \cite{Akers:2019gcv,Hayden:2021gno,Mori:2025mcd} for arguments for large tripartite entanglement and \cite{Harper:2021uuq,Harper:2022sky} for hints for large multi-partite entanglement in holography.}, we conjecture that this is indeed the case for holographic QEC codes.

$\vspace{-10mm}$
\section{Genuine multi-entropy in holography}
\label{sec:multiway}
\begin{figure}[t]
    \centering
    \begin{subfigure}{0.3\linewidth}
    \centering
    \includegraphics[width=\linewidth]{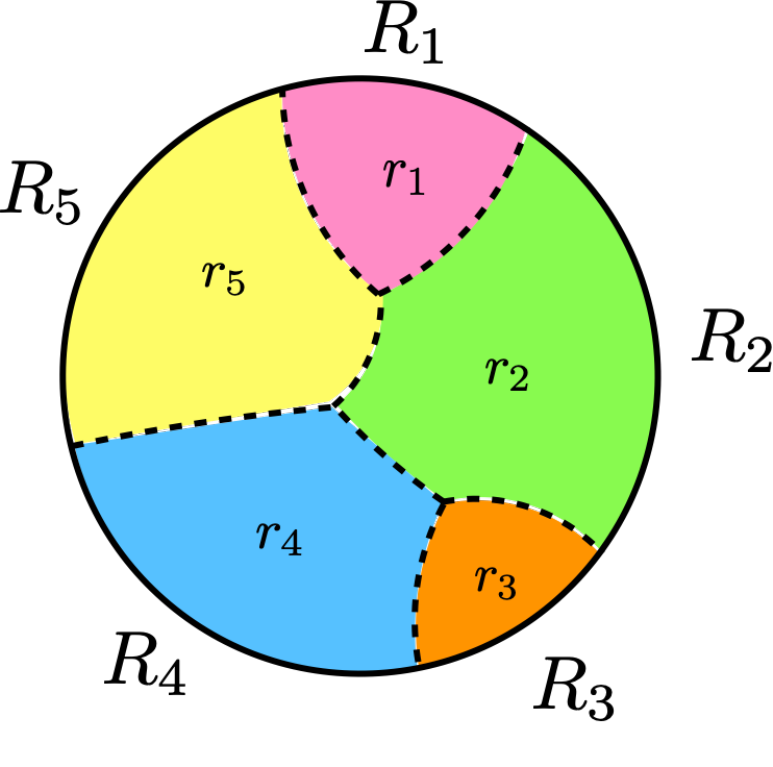}
    \vspace{-.93cm}
    \caption{}
    \label{fig:multiway}
    \end{subfigure}
    \hspace{1cm}
    \begin{subfigure}{0.3\linewidth}
    \centering
    \includegraphics[width=\linewidth]{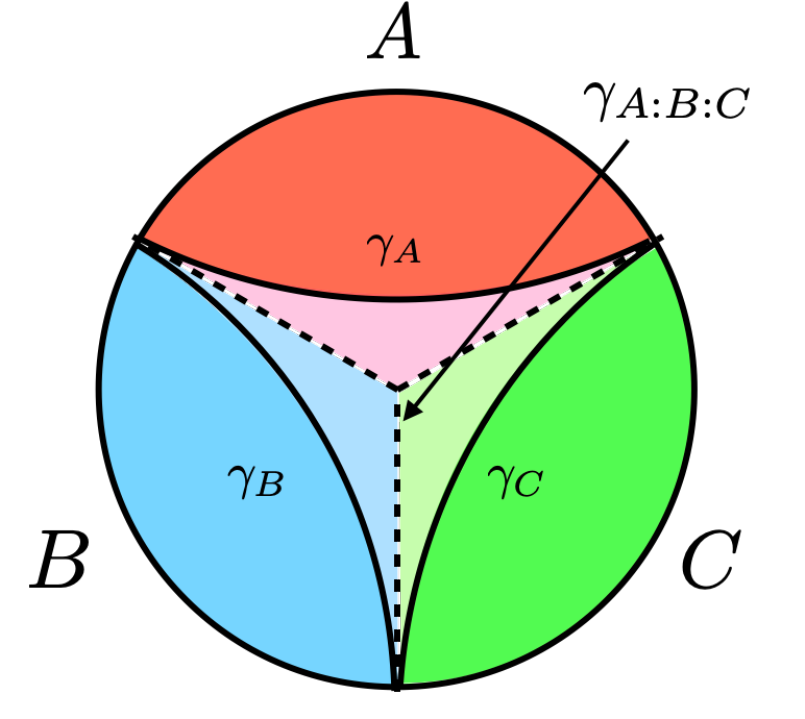}
    \caption{}
    \label{fig:tricut}
    \end{subfigure}
    \caption{\onehalfspacing (a) A generic 5-multiway cut. The area of the cut is the sum of the area of all dashed lines. Note that the intersection of the minimal surfaces is always trivalent. (b) The minimal surfaces relevant for the computation of $\GM[3]$.}
\end{figure}

Another piece of evidence for our claim on large multipartite entanglement comes from the structure of minimal surfaces in AdS/CFT. Consider a constant time slice $\Sigma$ of a $D$-dimensional AdS space. We work with $D=3$ below but we expect a similar argument holds true for generic $D$.
We divide the asymptotic boundary $\partial\Sigma$ into $\mathtt{q}\ge 3$ connected subregions $R_1,R_2,\cdots,R_\mathtt{q}$ and evaluate the genuine multi-entropy $\GM[q] (R_1:R_2:\cdots:R_\mathtt{q})$ for the boundary CFT state. As shown in Sec.~\ref{sec:intro}, $\GM[q]$ is given by a linear combination of the multi-entropies $\S[p]$ with $\mathtt{p}\le \mathtt{q}$. The idea is that $S^{(\mathtt{p})}$ has a proposed holographic dual \cite{Gadde:2022cqi}\footnote{See also \cite{Harper:2024ker}.} as the area of \emph{$\mathtt{p}$-multiway cuts} (divided by $4G_N$) on $\Sigma$. A $\mathtt{p}$-multiway cut is the minimum-area division of $\Sigma$ into $p$ subregions $r_i$ such that each $r_i$ contains the boundary region $R_i$, see Fig.~\ref{fig:multiway} for an example. As a result $\GM[q]$ can be determined solely from the bulk geometry. In what follows we will give examples (both analytically and numerically) that in general 
\begin{align}
\GM[q]= O\left(\frac{1}{G_N}\right) \qquad (\mbox{in holography}).
\end{align}
This signals that the boundary CFT states dual to $\Sigma$ have large $\mathtt{q}$-partite entanglement.

\begin{figure}[t]
    \centering
    \begin{subfigure}{0.45\linewidth}
    \includegraphics[width=\linewidth]{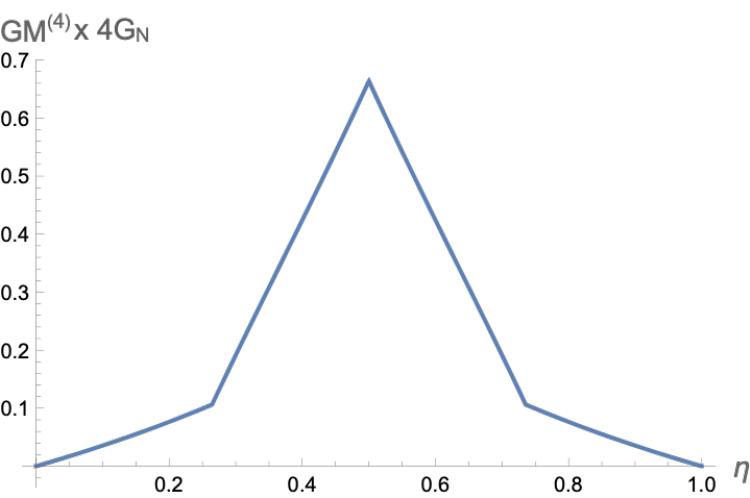}
    \caption{}
    \label{fig:GM4}
    \end{subfigure}
    \hspace{.5cm}
    \begin{subfigure}{0.47\linewidth}
    \includegraphics[width=\linewidth]{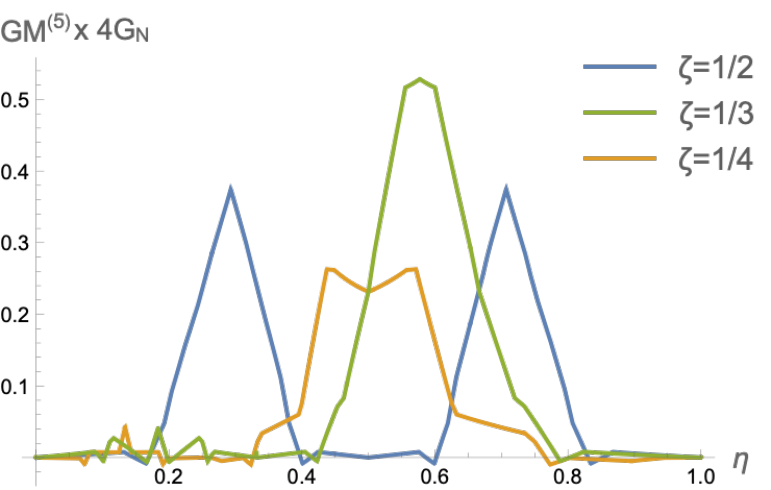}
    \caption{}
    \label{fig:GM5}
    \end{subfigure}
    \caption{\onehalfspacing (a) The holographic values of $\GM[4]$ in vacuum AdS$_3$, plotted against the conformal cross-ratio $\eta$ of the four points separating the boundary regions. We work with $a=1/3$ here. (b) The holographic values of $\GM[5]$ in vacuum AdS$_3$. There are five boundary separation points, which leaves us with two remaining degrees of freedom, which we take to be the the conformal cross ratios $(\eta,\zeta)$ of the first two points with respect to the remaining three. We work with $b=0$ here. We set $\ell_{\rm AdS}=1$ in the plots.}
\end{figure}

\begin{itemize}
\item $\mathtt{q}=3$ \\
  For tripartite divisions, the $\mathtt{q}=3$ genuine multi-entropy \eqref{GM3} is given by the following linear combination 
  \begin{align}
  \label{GM3geo}
    \GM[3](A:B:C) = \gamma_{A:B:C} - \frac{1}{2} (\gamma_A+\gamma_B+\gamma_C),
  \end{align}
  where $\gamma_{A:B:C}$ is the area of the triway cut (dashed lines in Fig.~\ref{fig:tricut}) and $\gamma_A,\gamma_B,\gamma_C$ are the area of the RT surfaces.
  It is not hard to see \footnote{An easy way to show this is to consider a splitting of the surface $\gamma_{A:B:C}$ into three 2-cuts each with $\frac{1}{2}$ weight and then make use of triangle inequality with respect to $\gamma_{A/B/C}$ \cite{Iizuka:2025ioc}.} that the quantity defined by \eqref{GM3geo} is semipositive and greater than zero if and only if the union of entanglement wedges of each individual regions does not equal the entire bulk region.\footnote{The same quantity \eqref{GM3geo} was used in Ref.~\cite{Akers:2024pgq} to argue for the existence of large tripartite entanglement in random tensor networks by showing that it lower bounds the \emph{Markov Gap} \cite{Hayden:2021gno} of the boundary state, another independent measure of tripartite entanglement.}, which is a defining feature of the holographic codes we considered. 
  
\item $\mathtt{q}=4$ \\
  For quadripartite divisions, $\GM[4]$ \eqref{GM4} comes with one free parameter $a$ :
  \begin{align}
    \label{GM4geo}
    &\GM[4](A:B:C:D) = \gamma_{A:B:C:D}
    -\frac{1}{3}(\gamma_{AB:C:D}+\gamma_{AC:B:D}+\cdots)\nonumber\\
    &\quad +a(\gamma_{AB}+\gamma_{AC}+\gamma_{AD})+(1/3-a)(\gamma_A+\gamma_B+\gamma_C+\gamma_D).
  \end{align}
  Albeit much more complicated, it is possible to prove that \cite{upcoming} the bulk geometric quantity corresponding to \eqref{GM4geo} is non-zero for $a\ge1/3$. Here we verify this fact numerically in Fig.~\ref{fig:GM4}.
  Combined with the results in Sec.~\ref{sec:QEC}, these are very strong evidence that the holographic CFT states have large quadripartite entanglement.
  
\item $\mathtt{q}\ge 5$ \\
  The above construction can be readily generalized into higher-partite divisions of the boundary. The expression for holographic genuine multi-entropies will involve increasingly more complicated linear combinations and more free parameters.
  In general we expect $\GM[q]$ to be nonzero whenever there is a bulk region that cannot be reconstructed from single boundary subregions, where the top multiway cut in the linear combination of $\GM[q]$ is non-trivial and distinct from all the lower $\mathtt{\tilde{q}}<\mathtt{q}$ cuts.
 This observation leads to our proposal that holographic CFT states must also contain large amounts of genuine $\mathtt{q}$-partite entanglement for any $\mathtt{q}\ge 5$. 
  For the record here we have evaluated $\GM[5]$ numerically and we verify that $\GM[5] \neq 0$ in general for generic 5-partite divisions of the boundary in Fig.~\ref{fig:GM5}. 
\end{itemize}

$\vspace{-10mm}$
\section{Discussion}

In this essay, we argued that the structure of holographic quantum error correction codes  and bulk minimal surfaces imply the existence of large higher-partite entanglement.
What can we say about the relative roles played by multipartite entanglements for different $\mathtt{q}$? A clue to answering this question comes from another important feature of holographic QEC codes. 
As indicated in Fig~\ref{fig:div4}, as we divide the boundary into finer and finer partitions, we lose access to more and more close-to-boundary regions in the bulk. Equivalently speaking, the construction of deeper bulk operators must rely more on higher-partite entanglement between finer boundary subregions than operators closer to the boundary\footnote{See Ref.~\cite{Pastawski:2015qua} for a similar discussion on multipartite entanglement and holographic QEC codes.}. In terms of a purely boundary point of view, this translates into the statement that higher $\mathtt{q}$-partite entanglements are increasingly more important in the IR, since it is known that radial direction in the bulk AdS can be interpreted as a RG flow of the boundary CFT \cite{Susskind:1998dq,deBoer:2000cz,Heemskerk:2010hk}.

Another evidence of the statement above pertains to minimal surfaces and multiway cuts. As shown in Sec.~\ref{sec:multiway}, the multiway-cut surfaces always lie within the region not reconstructable from any single boundary subsystems (the gray regions in Fig.~\ref{fig:div4}). Furthermore, on the boundary near the entangling surfaces, the contributions of all the multiway cuts cancel out. This also suggests a strong connection between higher-partite entanglement and bulk reconstructions of operators near the center of bulk.

$\vspace{-10mm}$
\section*{Acknowledgements}
S.L. would like to thank Ahmed Almheiri for insightful discussions and comments. The work of N.I. was supported in part by JSPS KAKENHI Grant Number 18K03619, MEXT KAKENHI Grant-in-Aid for Transformative Research Areas A “Extreme Universe” No. 21H05184, and NSTC of Taiwan Grant Number 114-2112-M-007-025-MY3.

\bibliographystyle{JHEP}
\bibliography{Refs}

\end{document}